\documentclass[12pt,prd,aps,superscriptaddress,preprintnumbers,amsmath,amssymb,nofootinbib]{revtex4-1}
\pdfoutput=1
\usepackage{amsmath,graphics,epsfig}
\usepackage{color,hyperref}
\usepackage{datetime}
\begin{document}

\title{A Heavy Scalar at the LHC from Vector Boson Fusion}
\author{Qiurong Mou}
\email{qiurongmou@163.com}
\affiliation{\small{Department of Physics, Chongqing University, Chongqing 401331, P. R. China}}
\author{Sibo Zheng}
\email{sibozheng.zju@gmail.com}
\affiliation{\small{Department of Physics, Chongqing University, Chongqing 401331, P. R. China}}
\date{July, 2018}

\begin{abstract}
A hypothetical scalar mixed with the standard model Higgs appears in few contexts of new physics.
This study addresses the question what mass range is in the reach of $14$ TeV LHC given different magnitudes of mixing angle $\alpha$,
where event simulations are based on production from vector boson fusion channel and decays into SM leptons through ${\rm WW}$ or ${\rm ZZ}$.
It indicates that heavy scalar mass up to $539$ GeV and $937$ GeV can be excluded by integrated luminosity of $300$ $\rm{fb}^{-1}$ and $3000$ $\rm{fb}^{-1}$ respectively 
for $\sin^{2}\alpha$ larger than $0.04$.
 
\end{abstract}

\maketitle
\setcounter{page}{0}
\thispagestyle{empty}

\newpage
\section{Introduction}
After the discovery of Higgs scalar $h$ at the LHC \cite{HiggsMass1, HiggsMass2} 
Standard Model (SM) as the effective field theory (EFT) of weak scale is established. 
While this EFT has not been violated at nowadays astrophysical and collider experiments,
it must be incomplete in the light of a few indirectly experimental as well as theoretic hints.
One of most robustly experimental hints arises from Plank and WIMP data \cite{1303.5076},
which suggests that there should be a new particle beyond SM serving as the thermal dark matter.
On the other hand, one of theoretic challenges is the need of some novel mechanism to stabilize the divergence involving Higgs mass.

In a few new physics models attempted to complete the EFT of SM such as SM with doublets and supersymmetry,
there usually exists a new scalar $H$ of the same spin, parity and quantum numbers with SM Higgs but with heavier mass.
Unless forbidden by some hidden symmetry, it generally mixes with the SM Higgs.
If so, such scalar may leave signatures at dark matter facilities 
which are in the reach of TeV mass scale.  See e.g. \cite{1609.03551, 1610.06292,1701.08134} for very recent studies on this subject.

Alternatively, $H$ can mix with the SM Higgs, and be examined at the LHC.
Due to mixing effect $H$ couplings to SM particles are similar to those of SM Higgs but with an universal scaling factor related to mixing angle smaller than unity.  
As a result, the diboson decay channels $H\rightarrow V_{i}V_{i}$ with $V_{i}$ referring $W$ or $Z$ boson dominate others for $H$ mass above $200$ GeV.
In this case, $H$ is mainly generated at the LHC through gluon gluon fusion (GGF) and vector boson fusion (VBF) channels
similar to the SM Higgs \cite{1101.0593}.
Early constraints \cite{0706.4311, 1303.3570,1303.1812} on the model parameters 
were obtained according to measurements on SM Higgs couplings, decay width and direct detection at the LHC.
Updated analyses based on the $8$ TeV LHC data \cite{1312.5353, 1405.3447, 1407.0558, 1407.6583, 1504.00936,1507.05930,1509.00389} 
can be found in \cite{1501.02234, 1502.01361,1503.01618,1505.05488, 1507.06158, 1601.07880}.

In this paper, we will employ the techniques reported in \cite{1504.00936}, 
and study the prospect for the discovery of $H$ at the $14$ TeV LHC through processes of VBF production and subsequent diboson decays to SM leptons final states
\footnote{The analysis here is more general than in the earlier version, which focused on the model interpretations of diboson excess reported in Ref. \cite{1506.00962}.}.
One reason is that although the GGF channel yields larger contribution to the production cross section than the VBF channel,
the contribution to SM background cross section arising from GGF process is also larger than VBF process.
Moreover, the ratio between GGF and VBF contribution to the production cross section declines from about $\sim 10$  to $\sim 2.5$ 
when  $m_{h_{2}}$ increases from $200$ GeV to $1$ TeV.

The paper is organized as follows. 
In Sec. II, we briefly discuss general parameterization of mixing effect in a model-independent way.
The key point is that only two model parameters appear in the following study of direct detection. 
In Sec. III, we address the production cross sections $\sigma(pp\rightarrow H+X)\times \text{Br}(H\rightarrow V_{i}V_{i}\rightarrow l\nu l\nu)$
from VBF channel at the LHC for $H$ mass above $200$ GeV.
Our main results are presented in Sec.IV, 
where we show the luminosities required for the $2\sigma$ exclusion and $5\sigma$ discovery.
Finally we conclude in Sec.V.

\section{Model}
Without mixing effect such as in the case of scalar dark matter model 
the mass squared matrix $\mathcal{M}^{2}$ for state vector $(H, h)$ reads as
\begin{eqnarray}{\label{decouple}}
\mathcal{M}^{2}=\left(%
\begin{array}{cc}
  m^{2}_{H} & 0 \\
  0& m^{2}_{h} \\
\end{array}%
\right)
\end{eqnarray}
where $m_{H}$ and $m_{h}$ denotes the mass of $H$ and $h$, respectively. 
There is little chance for direct detection on this kind of scalar at the LHC \cite{1601.06232}.
In contrast, in the case of mixing effect mass squared matrix $\mathcal{M}^{2}$ in Eq.(\ref{decouple}) should be replaced by,
\begin{eqnarray}{\label{nondecouplemass}}
\mathcal{M}^{2}=\left(%
\begin{array}{cc}
  m^{2}_{H} & \Delta m^{2} \\
  \Delta m^{2} & m^{2}_{h} \\
\end{array}%
\right)
\end{eqnarray}
where $\Delta m^{2}$ characterizes the mixing effect.

At present status only small mixing effect is allowed based on the LHC searches such as dijet, diphoton and four lepton signals.
The mass eigenvalues can be approximated to be \footnote{Note that the analytic rather than approximations here will be utilized for the numerical calculation in the next section.}
\begin{eqnarray}{\label{mass}}
m^{2}_{h_{2}}&\simeq& m^{2}_{H}+\frac{(\Delta m^{2})^{2}}{m^{2}_{H}-m^{2}_{h}}, \nonumber\\
m^{2}_{h_{1}}&\simeq& m^{2}_{h}-\frac{(\Delta m^{2})^{2}}{m^{2}_{H}-m^{2}_{h}},
\end{eqnarray}
together with their couplings to SM particles relative to SM Higgs
\begin{eqnarray}{\label{coupling}}
\frac{g^{2}_{h_{1}XX}}{g^{2}_{h_{\text{SM}}XX}}\simeq \cos^{2}\alpha,~~~~~~~~
\frac{g^{2}_{h_{2}XX}}{g^{2}_{h_{\text{SM}}XX}}\simeq \sin^{2}\alpha,
\end{eqnarray}
Here,  $X$ refers to the SM vector bosons and fermions,  
and the mixing angle $\alpha$ is given by
\begin{eqnarray}{\label{alpha}}
\tan(2\alpha)\simeq \frac{2\Delta m^{2}}{m^{2}_{H}-m^{2}_{h}}.
\end{eqnarray}
From Eq.(\ref{mass}) to Eq.(\ref{alpha}) one finds that 
the productions and decays of these two scalars are totally determined by heavier mass $m_{h_{2}}$ and mixing angle $\sin\alpha$
after identifying $h_{1}$ as the SM-like Higgs.
The magnitude of  $\sin^{2}\alpha$ has been up bounded to be less than $\sim 0.2$ at $95\%$ CL in the light of
precise measurement \cite{1303.3570,1303.1812} on the SM Higgs couplings at the $8$ TeV LHC,
and it will be improved to be of order $\sim 0.04$ at the future $14$ TeV LHC with designed integrated luminosity \cite{1310.8361}.

\section{Vector Boson Fusion}
In this section we address event simulation 
for the production cross section $\sigma(pp\rightarrow h_{2}+X)$ from VBF channel  
and branching ratios $\text{Br}(h_{2}\rightarrow V_{i}V_{i}\rightarrow l\nu l\nu)$ at the $14$ TeV LHC.
In particular, we use package FeynRules \cite{1310.1921} to generate model files prepared for MadGraph5 \cite{1405.0301},
which includes Pythia 6 \cite{0603175} for parton showering and hadronazition and the package Delphes 3 \cite{1307.6346} for fast detector simulation.

\subsection{Production Cross Section}
\begin{figure}
\centering
\includegraphics[width=8cm,height=8cm]{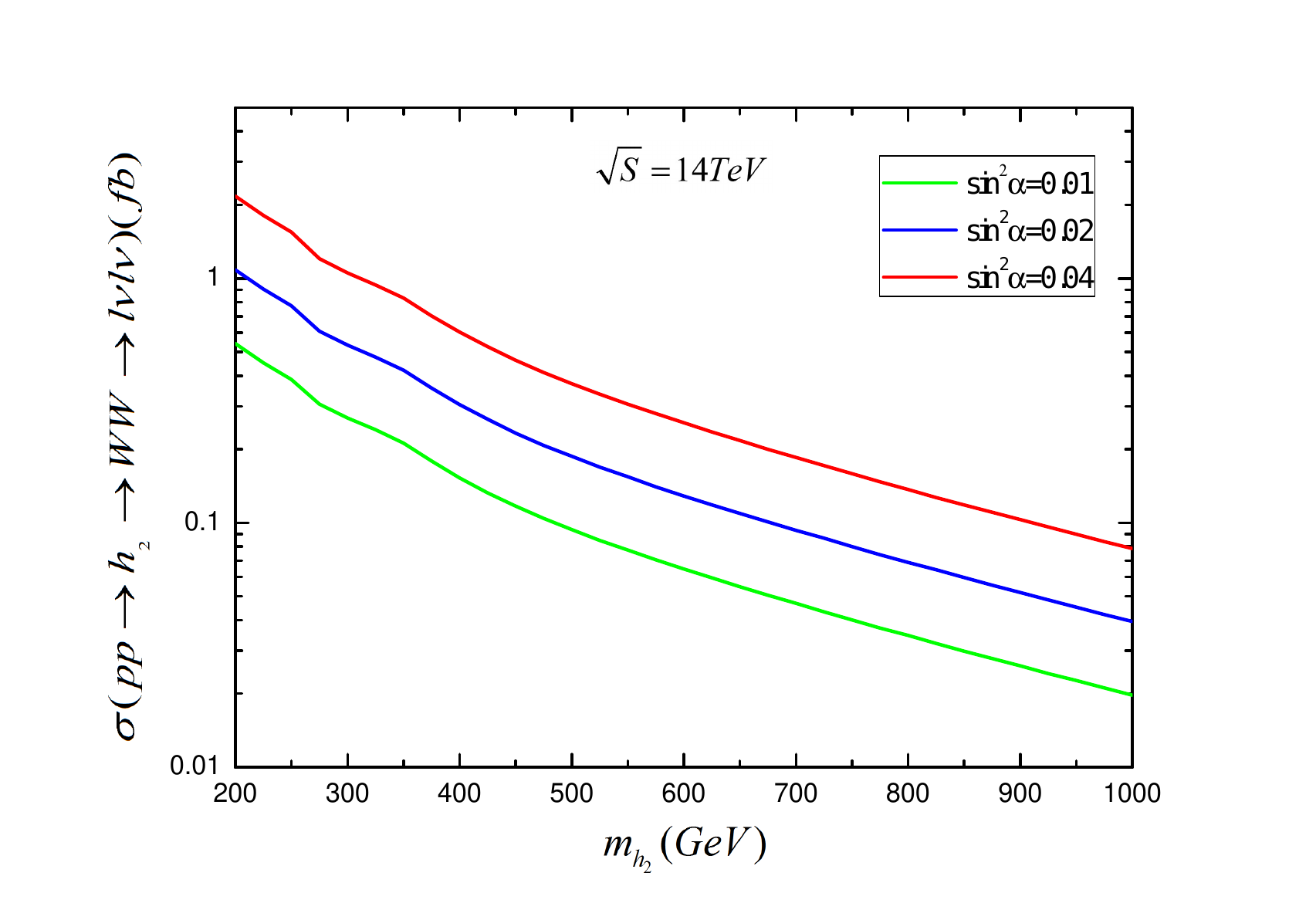}
\includegraphics[width=8cm,height=8cm]{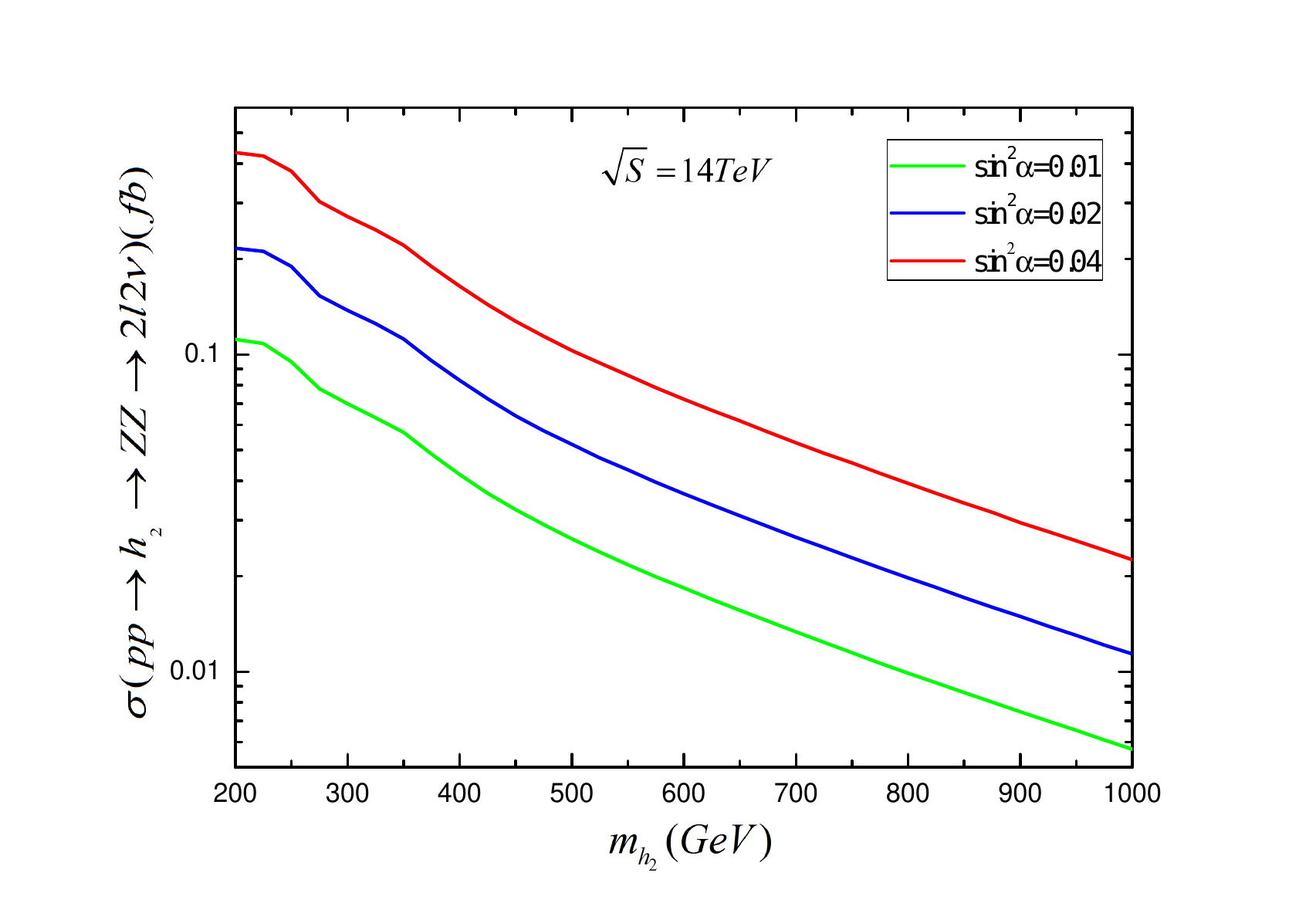}
\centering
 \caption{Production cross section 
 $\sigma(pp\rightarrow h_{2}+X)\times \text{Br}(h_{2}\rightarrow WW\rightarrow l\nu l\nu)$ (left) and
 $\sigma(pp\rightarrow h_{2}+X)\times \text{Br}(h_{2}\rightarrow ZZ\rightarrow 2l2\nu)$ (right) as function of $m_{h_{2}}$ 
 for different strengths of mixing effect $\sin^{2}\alpha=0.01$ (green), $0.02$ (blue) and $0.04$ (red), respectively.   }
\label{crosssections}
\end{figure}

We show in Fig.\ref{crosssections} the strengths of cross sections for  two different four-lepton final states,
where the dependence on the mixing angle can be understood as follows. 
Firstly, according to Eq.(\ref{coupling}) the VBF induced cross section $\sigma(pp\rightarrow h_{2})$  is proportional to $\sin^{2}\alpha$.
Secondly, with the definition on branching ratios $\text{Br}(h_{2}\rightarrow V_{i}V_{i})=\Gamma(h_{2}\rightarrow V_{i}V_{j})/\Gamma_{h_{2}}$, 
where $\Gamma_{h_{2}}=\Gamma(h_{2}\rightarrow \text{X}~\text{X})+\Gamma(h_{2}\rightarrow h_{1}h_{1})$ ($X$ is a SM fermion or vector boson),
$\text{Br}(h_{2}\rightarrow V_{i}V_{i})$ depends on the magnitude of $\Gamma(h_{2}\rightarrow h_{1}h_{1})$ relative to $\Gamma(h_{2}\rightarrow \text{X}~\text{X})$.
Unlike $\Gamma(h_{2}\rightarrow \text{X}~\text{X})$ which is determined by the mixing effects in quadratic term of scalar potential,  
$\Gamma(h_{2}\rightarrow h_{1}h_{1})$ is directly related to the cubic term in the scalar potential, which is model dependent.
For example, in the minimal supersymmetric standard model the ratio $\Gamma(h_{2}\rightarrow h_{1}h_{1})/\Gamma(h_{2}\rightarrow \text{X}~\text{X})$ is small for $m_{h_{2}}$ above $300$ GeV \cite{Gunion}, which implies that $\text{Br}(h_{2}\rightarrow V_{i}V_{i})$ mildly depends on the mixing angle. 
In contrast,  $\Gamma(h_{2}\rightarrow h_{1}h_{1})$ can be important in models such as extended Higgs doublet models,
where $\text{Br}(h_{2}\rightarrow V_{i}V_{i})$ will be related to parameters such as mixing angle, quadratic and cubic terms in the scalar potential.
For simplicity, we consider the case in which $\Gamma(h_{2}\rightarrow h_{1}h_{1})$ can be ignored.

\begin{figure}
\centering
\includegraphics[width=9cm,height=8cm]{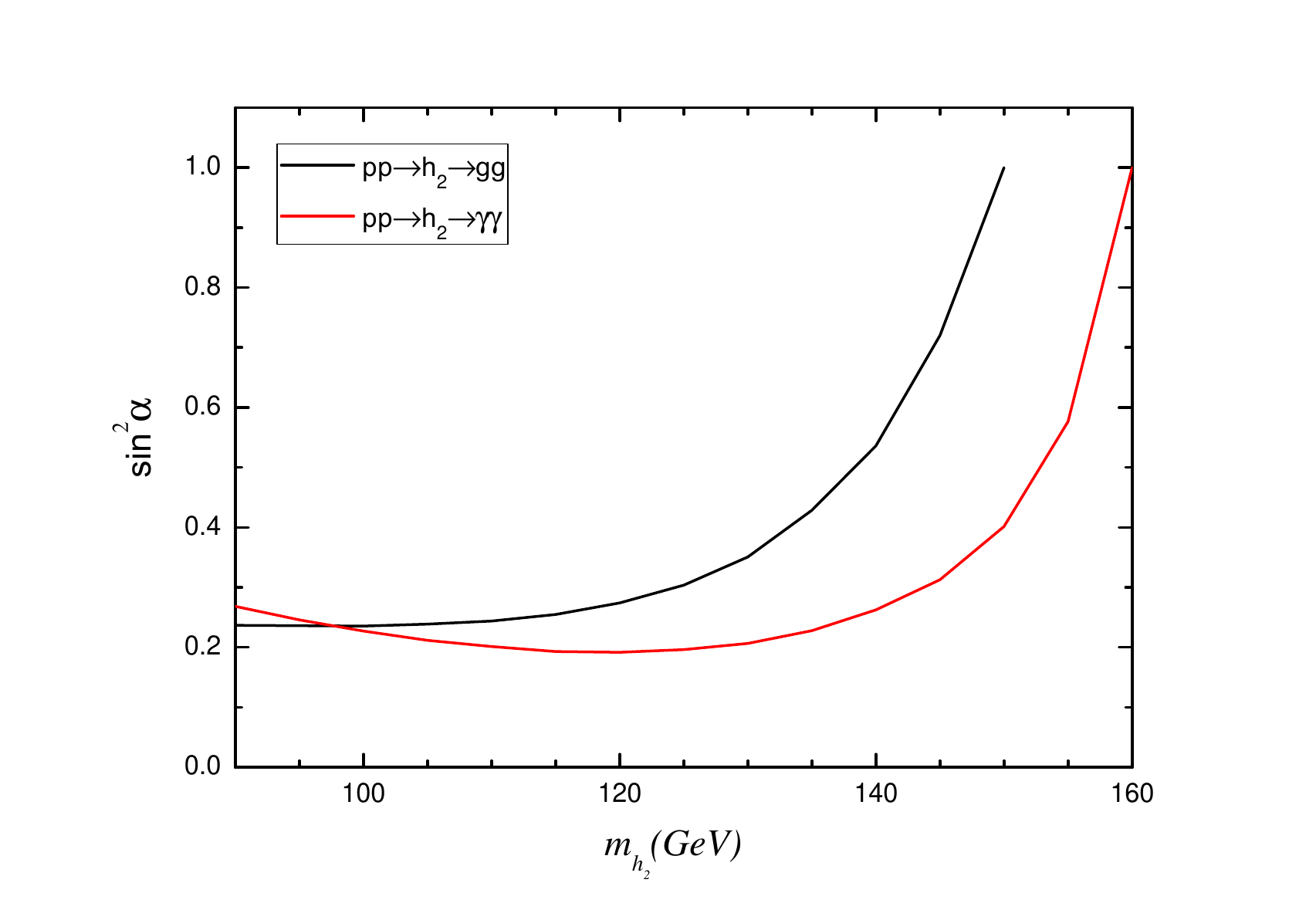}
\centering
 \caption{Direct constraint on mixing angle from decay modes $h_{2}\rightarrow\gamma\gamma$  \cite{twogamma} and
 $h_{2}\rightarrow\text{gg}$ \cite{dijet1, dijet2} at the 8 TeV LHC, which is verified to be much weaker than indirect constraint from precision tests on SM Higgs.}
\label{constraint}
\end{figure}

The parameter space composed of mixing angle and heavy scalar mass is subject to both direct and indirect constraints.
Current direct constraints include the $8$ TeV LHC bounds such as 
$\sigma_{\text{GGF+VBF}}\left(pp\rightarrow h_{2}\right)\times \text{Br}(h_{2}\rightarrow\text{gg})\leq 200$ fb \cite{dijet1, dijet2} and 
$\sigma_{\text{GGF+VBF}}\left(pp\rightarrow h_{2}\right)\times \text{Br}(h_{2}\rightarrow\gamma\gamma)\leq 0.5$ fb \cite{twogamma} in the mass region below $200$ GeV, as well as $\sigma_{\text{VBF}}\left(pp\rightarrow h_{2}\right)\times \text{Br}(h_{2}\rightarrow \text{ZZ})$ \cite{1507.05930}
and  $\sigma_{\text{VBF}}\left(pp\rightarrow h_{2}\right)\times \text{Br}(h_{2}\rightarrow \text{WW})$ \cite{1509.00389} in the mass region above $200$ GeV.
For illustration, we have shown in Fig.\ref{constraint} direct constraint on mixing angle in low mass region.
On the other hand, indirect constraints include precision measures on the Yukawa couplings of SM Higgs $h_1$ to SM fermions and vector bosons.
Global fits such as in \cite{1303.1812} implies that $\sin^{2}\alpha$ above $0.2$ has been excluded. 
Other indirect constrains arising from measurements on precision electroweak observables may also be useful to constraint the mixing angle.
In this sense, the constraint on mixing angle from indirect detection is much stronger than that from direct detection.

\subsection{Events Selection}
Let us now stress the event selections for the two different SM lepton final states from VBF channel.
The primary SM backgrounds to the first process $h_{2} \rightarrow WW\rightarrow l\nu l\nu$ 
include dilepton $+$ jets and QCD multi-jets. 
For simplicity, we consider the main contributions arising from dilepton plus jets channels, 
and adopt the cuts used by the CMS VBF analysis \cite{1504.00936} for event selection:
\begin{eqnarray}{\label{selection1}}
p_{T}^{l_{1}} > 20 ~{\rm GeV}, ~ p_{T}^{l_{2}}> 10 ~{\rm GeV}, 
~ p_{T}^{j_{1,2}} > 500 ~{\rm GeV}, &\,& \nonumber \\
|\eta_{e}| < 2.5, |\eta_{\mu}| < 2.4, ~|\Delta \eta_{jj}| > 3.5, &\,&~\nonumber\\
M_{ll}>12~{\rm GeV}, ~M_{jj} > 500~ {\rm GeV},~E^{{\rm miss}}_{{\rm T,Pr}} > 20~ {\rm GeV},~&\ &~
\end{eqnarray}
where ${p_{T}}^{l_{1(2)}}$ and ${p_{T}}^{j_{1(2)}}$ are the transverse momentum of the first (second) leading lepton $l=\{e,\mu\}$ and jet, respectively;
$\eta_{e(\mu)}$ is pseudo-rapidity of $e$($\mu$);
$\Delta \eta_{jj}$ and $M_{jj(ll)}$ is the rapidity difference and invariant mass of the two leading jets (leptons), respectively.
Parameter $E^{{\rm miss}}_{{\rm T,Pr}}$ is defined as 
\begin{equation}{\label{Emiss}}
E^{{\rm miss}}_{{\rm T,Pr}}=
 \left\{
\begin{array}{lcl}
E^{{\rm miss}}_{T}\cdot \cos(\Delta\Phi),~~~~~~~~~~~\Delta\Phi(p_{T}, E_{T}^{\rm{miss}})<\pi/2,\\
\overrightarrow{E}^{{\rm miss}}_{T},~~~~~~~~~~~~~~~~~~~~~~~~\Delta\Phi(p_{T}, E_{T}^{\rm{miss}})>\pi/2.
\end{array} \right. 
\end{equation}
with $\Delta\Phi(p_{T}, E_{T}^{\rm{miss}})$ referring to the azimuthal angle between the dilepton transverse momentum and
$\overrightarrow{E}_{T}^{\rm{miss}}$.
Any event with an additional jet with $p_T > 30 $ GeV is rejected.
We refer the reader to Ref. \cite{1504.00936} for more details.

For the second process $h_{2} \rightarrow ZZ \rightarrow 2l2\nu$ 
we consider the main contributions arising from electron pair $+$ jets $+$ $E_{T}^{\rm{miss}}$
and muon pair $+$ jets $+$ $E_{T}^{\rm{miss}}$.
Cuts \cite{1504.00936} for event selection in this channel are given by
\begin{eqnarray}{\label{selection2}}
p_{T}^{l1(2)} > 20~{\rm GeV},~p_{T}^{ll} > 55~{\rm GeV},~p_{T}^{j} > 30~{\rm GeV}, , &\,& \nonumber \\
 |\eta_{j}| <2.5,~|\Delta\eta_{jj}|> 4, ~M_{jj} > 500~ {\rm GeV}, &\,& ~\nonumber\\
60~{\rm GeV} < M_{ll}< 120~{\rm GeV},~E_{T}^{\rm{miss}}\geq 70~{\rm GeV},~&\,&
\end{eqnarray}
where $p_{T}^{ll}$ denotes the $p_{T}$ of the dilepton system.
Any event which includes the third lepton with $p_T > 20 $ GeV is rejected in order to suppress SM $WZ$ background.
More details can be also found in Ref. \cite{1504.00936}.

Cuts in Eq.(\ref{selection1}) and Eq.(\ref{selection2}) will be applied to the 14 TeV LHC simulations for conservation,
the validity of which is guaranteed by the following facts. 
At first, there is little difference between the 8 TeV LHC and 14 TeV LHC except the collision energy, 
which means the cut on the pseudo-rapidity of the first two leading jets should remain unchanged. 
Second, the kinetic distribution of the signal events and the main SM backgrounds have similar changing trends
when one modifies these cuts.
Take the representative mass $m_{h_{2}}=600$ GeV for example.
The effects on the ratio of signal over background events $S/B$ are less than two times due to variations on the cuts in Eq.(\ref{selection1}) and Eq.(\ref{selection2}).
See Table \ref{ratio} for details.

\begin{table}
\footnotesize
\begin{center}
\begin{tabular}{|c|c|c|c|}
 \hline
 $p_{T}^{l1}>\{10, 20, 40\}$ GeV & $p_{T}^{j1}>\{400, 500, 600\}$ GeV &  $\Delta\eta_{jj} >\{3, 3.5, 4.0\}$ & 
$M_{jj} >\{400, 500, 600\}$ GeV    \\
 $\{0.98,1, 1.02\}$ & $\{1,1,0.99\}$  & $\{0.79,1,1.38\}$  &  $\{0.82,1,1.36\}$ \\
\hline\hline
 $p_{T}^{e}>\{10, 20, 40\}$ GeV & $p_{T}^{j1}>\{20, 30, 40\}$ GeV &  $\Delta\eta_{jj} >\{3.5, 4, 4.5\}$ & 
$M_{jj} >\{400, 500, 600\}$ GeV    \\
 $\{1,1,0.99\}$ & $\{1,1,0.98\}$  & $\{1.51,1,0.7\}$  &  $\{1.13, 1, 0.90\}$ \\
 \hline
 \end{tabular}
 \caption{Effects on the ratio $S/B$ due to variations on the cuts in Eq.(\ref{selection1}) (top) and Eq.(\ref{selection2}) (bottom) 
 for benchmark mass $m_{h_{2}}= 600$ GeV at $14$ TeV LHC.}
 \label{ratio}
\end{center}
\end{table}

\begin{figure}
\centering
\includegraphics[width=8cm,height=8cm]{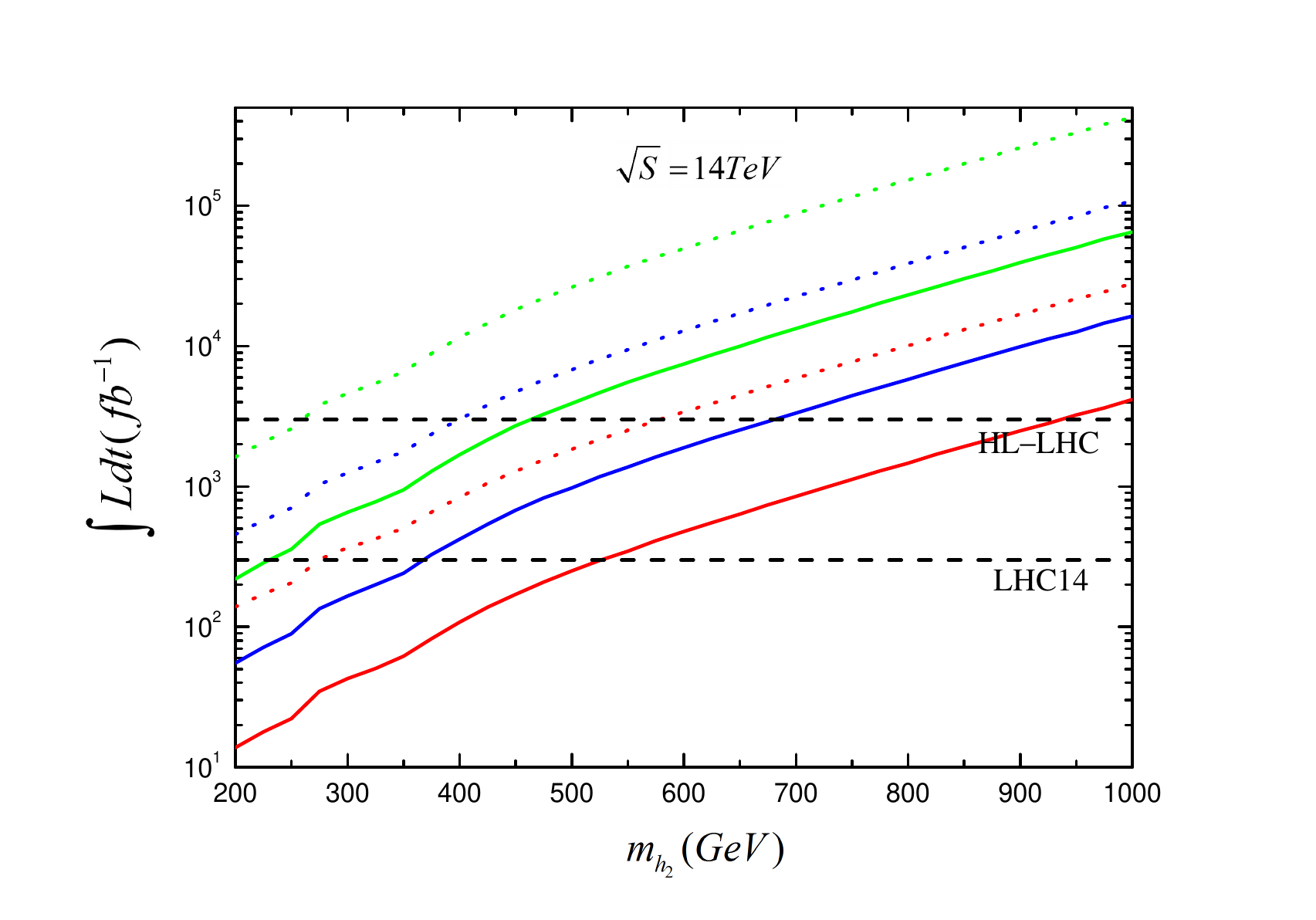}
\includegraphics[width=8cm,height=8cm]{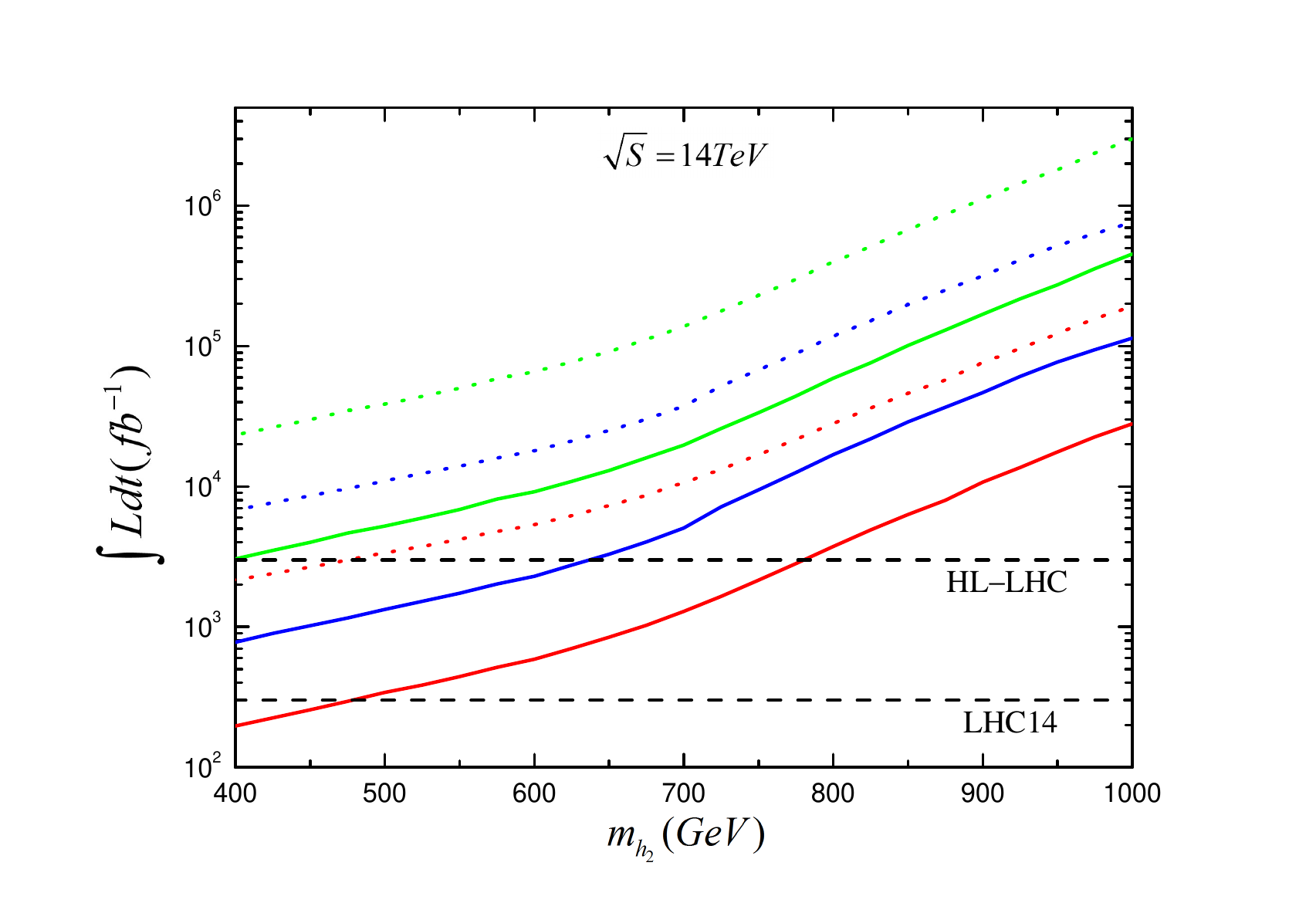}
\centering
 \caption{The integrated luminosity needed for the exclusion determined by $S/\sqrt{B}=1.96$ (solid) 
 and $5\sigma$ discovery determined by $S/\sqrt{S + B}=5$ (dotted) at the 14 TeV LHC, respectively.
 Here, the left and right plot corresponds to the left and right plot of Fig.\ref{crosssections}, respectively, 
 where the meaning of colors is the same as in Fig.\ref{crosssections}. 
The two horizontal lines correspond to integrated luminosity of $300$ $\rm{fb}^{-1}$ and $3000$ $\rm{fb}^{-1}$. }
\label{results}
\end{figure}

\section{Results}
Following the definition $S/\sqrt{B}$ and $S/\sqrt{S+B}$ about significance for exclusion and discovery, respectively, 
we show the values of $L$ needed for exclusion and discovery at the 14 TeV LHC in Fig.\ref{results}.
Systematic uncertainties are neglected in both the signal and the background simulations.

The left plot therein address the decay $h_{2}\rightarrow WW\rightarrow l\nu l\nu$.
In the left one we observe that for $L=300$ $\rm{fb}^{-1}$
$h_2$ mass up to $\{270, 368, 539\}$ GeV can be excluded via final state $l\nu l\nu$ for $\sin^{2}\alpha=\{0.01,0.02,0.04\}$, respectively;
and the discovery limit approaches to $275$ GeV for $\sin^{2}\alpha=0.04$.
Furthermore, for HL-LHC with $L=3000$ $\rm{fb}^{-1}$ \cite{1310.8361}
$h_2$ mass up to $\{459, 675, 937\}$ GeV can be excluded via this channel for $\sin^{2}\alpha=\{0.01,0.02,0.04\}$, respectively;
and the discovery limits reach $\{267, 401, 583\}$ GeV for $\sin^{2}\alpha=\{0.01,0.02,0.04\}$, respectively.

The right plot in Fig.\ref{results} addresses the decay $h_{2}\rightarrow ZZ\rightarrow 2l2\nu$. 
It shows that for $L=300$ $\rm{fb}^{-1}$
$h_2$ mass up to $475$ GeV can be excluded through final state $2l2\nu$ for $\sin^{2}\alpha=0.04$.
Moreover, for the HL-LHC $h_2$ mass up to $\{400, 640, 790\}$ GeV can be excluded via the same channel for $\sin^{2}\alpha=\{0.01, 0.02,0.04\}$, respectively;
and the discovery limit reaches $477$ GeV for $\sin^{2}\alpha=0.04$.
The exclusion limits via the $ZZ$ decay are relatively weaker in comparison with the $WW$  decay.

\section{Conclusions}
Hypothetical scalar similar to the SM Higgs often appears in a complete model of quantum field theory.
This work is devoted to study a heavy scalar mixed with SM Higgs at the $14$ TeV LHC through VBF channel. 
We have simulated events arising from diboson decays such as $h_{2}\rightarrow WW\rightarrow l\nu l\nu$ and $h_{2}\rightarrow ZZ\rightarrow2l2\nu$,
where both exclusion and discovery limits are revealed according to different magnitudes of mixing effect.
Our study demonstrates that such type of heavy scalar with mass up to $539$ GeV and $937$ GeV can be excluded by the $14$ TeV LHC with $L=300$ $\rm{fb}^{-1}$ and $3000$ $\rm{fb}^{-1}$ respectively 
for $\sin^{2}\alpha$ larger than $0.04$.\\

$\mathbf{Conflict~of~Interest}$.
The authors do not have a direct financial relation with any commercial identity mentioned in the paper that might lead to a conflict of interests for any of the authors.

$\mathbf{Acknowledgments}$.
This work is supported in part by the National Natural Science Foundation of China under Grant No.11775039.


\begin{thebibliography}{99}
\bibitem{HiggsMass1}
G.~Aad {\it et al.} [ATLAS Collaboration],
Phys.\ Lett.\ B {\bf 716}, 1 (2012), [arXiv:1207.7214 [hep-ex]].

\bibitem{HiggsMass2}
 S.~Chatrchyan {\it et al.} [CMS Collaboration],
Phys.\ Lett.\ B {\bf 716}, 30 (2012),
[arXiv:1207.7235 [hep-ex]].

\bibitem{1303.5076}
Planck Collaboration,
 ``Planck 2013 results. XVI. Cosmological parameters,''
 Astron.\ Astrophys.\  {\bf 571}, A16 (2014),
 arXiv:1303.5076 [astro-ph.CO].

\bibitem{1609.03551}
 X.~G.~He and J.~Tandean,
JHEP {\bf 1612}, 074 (2016),
[arXiv:1609.03551 [hep-ph]].

\bibitem{1610.06292}
  H.~Wu and S.~Zheng,
JHEP {\bf 1703}, 142 (2017),
[arXiv:1610.06292 [hep-ph]].

\bibitem{1701.08134}
J.~A.~Casas, D.~G.~Cerdeño, J.~M.~Moreno and J.~Quilis,
JHEP {\bf 1705}, 036 (2017),
[arXiv:1701.08134 [hep-ph]].


\bibitem{1101.0593}
 S.~Dittmaier {\it et al.} [LHC Higgs Cross Section Working Group],
[arXiv:1101.0593 [hep-ph]].



\bibitem{0706.4311}
  V.~Barger, P.~Langacker, M.~McCaskey, M.~J.~Ramsey-Musolf and G.~Shaughnessy,
Phys.\ Rev.\ D {\bf 77}, 035005 (2008),
[arXiv:0706.4311 [hep-ph]].

\bibitem{1303.3570}
P. P. Giardino, K. Kannike, I. Masina, M. Raidal, and A. Strumia,
JHEP 1405 (2014) 046, [arXiv:1303.3570 [hep-ph]].

\bibitem{1303.1812}
A. Falkowski, F. Riva, and A. Urbano, 
JHEP 1311 (2013) 111, [arXiv:1303.1812 [hep-ph]].



\bibitem{1312.5353}
  S.~Chatrchyan {\it et al.} [CMS Collaboration],
Phys.\ Rev.\ D {\bf 89}, no. 9, 092007 (2014).
[arXiv:1312.5353 [hep-ex]].

\bibitem{1405.3447}
 V.~Khachatryan {\it et al.} [CMS Collaboration],
JHEP {\bf 1408}, 174 (2014),
[arXiv:1405.3447 [hep-ex]].


\bibitem{1407.0558}
 V.~Khachatryan {\it et al.} [CMS Collaboration],
Eur.\ Phys.\ J.\ C {\bf 74}, no. 10, 3076 (2014).
[arXiv:1407.0558 [hep-ex]].


\bibitem{1407.6583}
G.~Aad {\it et al.} [ATLAS Collaboration],
Phys.\ Rev.\ Lett.\  {\bf 113}, no. 17, 171801 (2014).
[arXiv:1407.6583 [hep-ex]]. 

\bibitem{1504.00936}
V.~Khachatryan {\it et al.} [CMS Collaboration],
JHEP {\bf 1510}, 144 (2015),
[arXiv:1504.00936 [hep-ex]].

\bibitem{1507.05930}
 G.~Aad {\it et al.} [ATLAS Collaboration],
Eur.\ Phys.\ J.\ C {\bf 76}, no. 1, 45 (2016),
[arXiv:1507.05930 [hep-ex]].


\bibitem{1509.00389}
 G.~Aad {\it et al.} [ATLAS Collaboration],
JHEP {\bf 1601}, 032 (2016),
[arXiv:1509.00389 [hep-ex]].



\bibitem{1501.02234}
  T.~Robens and T.~Stefaniak,
Eur.\ Phys.\ J.\ C {\bf 75}, 104 (2015),
[arXiv:1501.02234 [hep-ph]].

\bibitem{1502.01361}
  A.~Falkowski, C.~Gross and O.~Lebedev,
JHEP {\bf 1505}, 057 (2015),
[arXiv:1502.01361 [hep-ph]].


\bibitem{1503.01618}
S.~I.~Godunov, A.~N.~Rozanov, M.~I.~Vysotsky and E.~V.~Zhemchugov,
Eur.\ Phys.\ J.\ C {\bf 76}, 1 (2016),
[arXiv:1503.01618 [hep-ph]].


\bibitem{1505.05488}
 D.~Buttazzo, F.~Sala and A.~Tesi,
JHEP {\bf 1511}, 158 (2015),
[arXiv:1505.05488 [hep-ph]].


\bibitem{1507.06158}
 K.~Cheung, P.~Ko, J.~S.~Lee and P.~Y.~Tseng,
JHEP {\bf 1510}, 057 (2015),
[arXiv:1507.06158 [hep-ph]].


\bibitem{1601.07880}
T.~Robens and T.~Stefaniak,
Eur.\ Phys.\ J.\ C {\bf 76}, no. 5, 268 (2016),
[arXiv:1601.07880 [hep-ph]].



\bibitem{1506.00962}
G.~Aad {\it et al.} [ATLAS Collaboration],
JHEP {\bf 1512}, 055 (2015),
[arXiv:1506.00962 [hep-ex]].


\bibitem{1601.06232}
H.~Han, J.~M.~Yang, Y.~Zhang and S.~Zheng,
Phys.\ Lett.\ B {\bf 756}, 109 (2016),
[arXiv:1601.06232 [hep-ph]].



\bibitem{1310.8361}
 S.~Dawson {\it et al.},
[arXiv:1310.8361 [hep-ex]].


\bibitem{1310.1921}
  A.~Alloul,  {\it et al.},
  Comput.\ Phys.\ Commun.\  {\bf 185}, 2250 (2014)
  [arXiv:1310.1921 [hep-ph]].

 \bibitem{1405.0301}
 J.~Alwall {\it et al.},
JHEP {\bf 1407}, 079 (2014),
[arXiv:1405.0301 [hep-ph]].


\bibitem{0603175}
 T.~Sjostrand, S.~Mrenna and P.~Z.~Skands,
JHEP {\bf 0605}, 026 (2006)
[hep-ph/0603175].

\bibitem{1307.6346}
J.~de Favereau {\it et al.} [DELPHES 3 Collaboration],
JHEP {\bf 1402}, 057 (2014),
[arXiv:1307.6346 [hep-ex]].


\bibitem{Gunion}
J.~F.~Gunion, H.~E.~Haber, G.~L.~Kane and S.~Dawson,
Front.\ Phys.\  {\bf 80}, 1 (2000).


\bibitem{dijet1}
G.~Aad {\it et al.} [ATLAS Collaboration],
Phys.\ Rev.\ D {\bf 91}, no. 5, 052007 (2015),
[arXiv:1407.1376 [hep-ex]].

\bibitem{dijet2}
V.~Khachatryan {\it et al.} [CMS Collaboration],
Phys.\ Rev.\ D {\bf 91}, no. 5, 052009 (2015),
[arXiv:1501.04198 [hep-ex]].


\bibitem{twogamma} 
G.~Aad {\it et al.} [ATLAS Collaboration],
[arXiv:1504.05511 [hep-ex]].







\end{thebibliography}
\end{document}